\documentclass[preprint,showpacs,preprintnumbers,amsmath,amssymb]{revtex4}

\usepackage{graphicx}
\usepackage{dcolumn}
\usepackage{bm}

\begin{document}

\preprint{APS/123-QED}

\title{$\Theta^+$: Another Explanation and Prediction}

\author{Tadafumi Kishimoto and Toru Sato}
\affiliation{%
Department Physics, Osaka University, 
Toyonaka, Osaka, 560-0043, Japan
}%


\date{\today}

\begin{abstract}
Recently the so-called $\Theta^+$ resonance has been reported first 
from SPring8\cite{nakano} and many following experiments showed 
apparent confirmation of the state.   Since $\Theta^+$ exclusively 
decays into either $K^+ n$ or $K^0 p$, it is explained as the 
predicted pentaquark state which includes $ u u d d \bar{s}$ 
quarks.  However, one yet has to obtain consistent picture of 
$\Theta^+$ and its quantum numbers.  We try to explain 
$\Theta^+$ in a conventional picture and show that such picture leads 
to new predictions on kaon and pion system.  
\end{abstract}

\pacs{13.75.Lb, 13.60.Le, 13.60.Rj}
\maketitle


In this letter we would like to present some remarks on the 
recently observed new exotic particle $\Theta^+$.   We show that our 
picture leads to further exotic prediction which experiments have likely 
been missing so far.   First let us briefly summarize what have been 
observed and can be assumed.   $\Theta^+$ was observed as a sharp peak at 
around 1540 MeV in the invariant mass spectrum of both $K^+ n$ and 
$K^0 p$ \cite{nakano, diana, clas,saphier}.  
The width is probably narrower than 10 MeV.  Its 
spin parity is $1/2^+$ according to original prediction by Diakonov 
\cite{Diakonov}.   Another property is that $\Theta^+$ seems to be 
observed only in charge +1 state which 
indicates that its isospin is 0.   Although no assignment on spin, parity 
and isospin have been given experimentally, it would be worth and 
interesting to consider what would be derived from the 
existence of $\Theta^+$ with such properties.  

These properties suggests us a conventional picture of $\Theta^+$ as 
a bound state of $K \pi N$.  The masses of the three constituent 
particles are 1568.2 MeV for $K^+ \pi^0 n$, 1570.9 MeV for $K^0 \pi^0 p$, and 
1576.8 MeV for $K^0 \pi^+ n$, respectively.  The $\Theta^+$ is at 
roughly 30 MeV below the threshold.  The binding energy of 30 MeV is 
quite suggestive to consider $\Theta^+$ as a bound state.  It is not 
extraordinary since $\sim$ 10 MeV par particle is typical in nuclei.   
Question is whether the interactions among $K , \ \pi$ and $N$ could 
realize the bound state.  
We assume that $\Theta^+$ has $J^\pi = 1/2^+$ and $I=0$ which 
are then quantum numbers of the $K \pi N$ system.   It is natural 
to consider that the three particles are in an s-wave to realize the 
lowest energy state.  Then $J^\pi$ of the $K \pi N$ system is equal to 
$1/2^+$.    Accordingly, spin, parity and isospin of any two particle 
subsystems are uniquely determined.  They are shown in table 1.   
\begin{table} 
\caption{\label{tab:table1}Spin, parity and isospin of two particle 
subsystems.  } 
\begin{ruledtabular} 
\begin{tabular}{c|c|c|c|c} 
& $\Theta^+ \ (K \pi N$) & $\pi N$ & $K N$ & $K \pi$ \\ 
\hline 
$J\pi$ & $1/2^+$ & $1/2^-$ & $1/2^-$ & $0^+$\\ 
$I$ & 0 & 1/2 & 1 & 1/2\\ 
\end{tabular} 
\end{ruledtabular} 
\end{table} 
The nice feature of the three body bound state is that the narrow 
width of $\Theta^+$ can be naturally explained.  Since $\Theta^+$ 
decays into $K N$, a pion has to be absorbed either in a nucleon or 
a kaon to decay.   The pion cannot be absorbed in the kaon because 
$0^+$ of the $K \pi$ system does not match $0^-$ of the kaon.  If 
one allows the $K \pi$ system to have relative angular momentum, the 
spin-parity can be $1^-, \ 2^+, \ 3^-$, and so on.  Therefore this 
mismatch cannot be resolved even though non-zero relative angular 
momentum is introduced.    The pion can be absorbed in the nucleon 
only when relative angular momentum of the $\pi N$ system be 
excited from s-wave to p-wave to make $J^\pi=1/2^+$.  
This excitation requires the kaon be excited in relative p-wave 
with respect to the $\pi N$ system simultaneously.  The decay can 
thus take place through weak mixing of p-wave.  

Such an admixture is seen in the deuteron.  It has a d-wave 
component which is an-order-of magnitude smaller than the dominant 
s-wave one.  It is due to the tensor interaction between nucleons.  
In the deuteron case total spin 1 allows the two nucleon system to 
have both s- and d-wave components simultaneously.  However, in the 
$\Theta^+$ case, introduction of p-wave in the $\pi N$ system is 
possible only when the kaon is also excited in p-wave with respect 
to $\pi N$ system.  Thus it is strongly suppressed.   Since $\Theta^+$ 
is a $K \pi N$ bound system in our view, it is an object much more 
extended than the typical strong interaction range.  The excitation to 
p-wave can take place only when particles are within the interaction 
range.  Thus the decay is further suppressed.  
Later we give an order-of-magnitude estimate of the width.

We calculated the $K \pi N$ system based on two body interactions for 
all three channels shown in table 1.   The $K N$ and $\pi N$ 
scattering have been studied and they are summarized in phase shift 
analysis \cite{arndt-pin, hashimoto, arndt-kn}.  The $K \pi$ 
interaction was derived from the reaction to produce the two 
particles simultaneously since both kaon and pion are unstable 
particles \cite{estabrooks}.  
In a calculation we used separable potential to reproduce the 
available phase shifts at low energy region.   We were not able to 
find any bound state.  We recognized similar attempt to explain 
$\Theta^+$ as a bound state of the three particles where, however, 
no bound state was demonstrated to exist \cite{7quark, Estrada}.   
The $\pi N$ 
channel (I=1/2) and $K \pi$ channel (I=1/2) are weakly attractive and 
$K N$ channel (I=1) is weakly repulsive.   The attractive 
interactions are too weak to realize the bound state.   The 
scattering lengths are $\sim 0.18 m_\pi^{-1}$ for $\pi N$(I=1/2) 
\cite{pin},  $\sim -0.21 m_\pi^{-1}$ for $K N$(I=1) \cite{kn} and 
$\sim 0.33 m_\pi^{-1}$ for $K \pi$(I=1/2) \cite{estabrooks}.  
These are roughly an-order-of magnitude smaller than that of 
nucleon-nucleon interaction.   We think that it is difficult to 
reproduce the 30 MeV binding energy with two body interactions 
currently available.  

In the present calculation we used two body interactions which 
explain low energy scattering data.  
Extraction of two body interaction at very low energy region is 
difficult if an unstable particle is relevant.  It is particularly 
difficult when both particles are unstable since assumed production 
mechanism of the two particles dominantly determines the interaction.  
Thus we think $K \pi$ interaction has much more room to be modified 
than that of $K N$ and $\pi N$.  We thus considered the case that two 
body interactions for $K N$ and $\pi N$ are fixed and that of $K \pi$ 
is set free.

We think that two body $K \pi$ interaction has to be very strong 
in order to reproduce binding energy of 30 MeV.   It is so strong 
that the $K \pi$ system has a bound state.   As far as we take 
into account two-body interactions in the $K \pi N$ system, there 
will be no bound state, if the $K \pi$ system has no bound state.    
Therefore, the idea of $\Theta^+$ as the $K \pi N$ bound state 
strongly suggests a possible existence of the bound state in the 
$K \pi$ system.  In other words a search for a bound state in the 
$K \pi$ system will provide us a key to answer the question on the 
nature of $\Theta^+$.   Let us call this presumed particle $X$ 
which yet has to be searched for.  

We do not give a prediction of binding energy of $X$ since it is 
already constrained in a region which is narrow enough for the 
experimental search.  If $X$ exists as a bound state, its binding 
energy has to be within a range of 0-30 MeV otherwise $\Theta^+$ 
decays into a $X$ and a nucleon.   Its spin parity is $0^+$ 
and $I=1/2$.  Since the state consists of only mesons, the charge 
conjugation leads to the existence of both $K \pi$ and $\bar{K} \pi$ 
systems.   Therefore $X^+$ and $X^0$ and their antiparticles $X^-$ 
and $\bar{X^0}$ exist.  
The prediction of $X$ as the $K \pi$ bound state is very exotic and hard to 
believe although we would like to point out that such possibility has 
not been completely ruled out.   We also would like to present an 
experiment to search for the state or to rule out this possibility.

One wonders that even though $X$ may possibly exist why it has 
escaped our experimental study.  Let us start our discussion 
on decay properties of $X$.  A particle with strangeness that is 
lighter than $X$ must be a kaon.  Since emission of a 
pion in addition to a kaon is energetically forbidden, no 
decay by strong interaction can take place.   On the other 
hand, $X$ can decay by the electromagnetic interaction.  
Particles possibly present with a kaon in the final state are 
$\gamma$'s and $e^+ e^-$ pairs.  The $X(0^+) \rightarrow K(0^-)$ 
transition constrains particles in the final state.  The $X 
\rightarrow K \gamma$ decay is forbidden because of angular 
momentum conservation.  The $X \rightarrow K e^+ e^-$ decay is 
forbidden because the electromagnetic vector current doesn't 
have an axial charge.  Therefore the $X \rightarrow K 
\gamma \gamma$ decay is the lowest order decay mode. 
Lifetime is then expected to be that of $\pi^0$.   
This decay mode makes it very difficult to 
identify $X$ by experiments.   Two $\gamma$'s from the $X$ decay 
make no peak in the invariant mass distribution.   Usually 
m$_{\pi^0}$ region is selected in the invariant mass distribution 
of two $\gamma$'s to make further hadron spectroscopy and this 
procedure leaves no chance to search for $X$.   Detection of 
$\gamma$ rays is usually difficult since $\gamma$-ray detectors 
are subject to backgrounds and loss of signal due to energy escape 
from electromagnetic shower in the detectors.  

One has to search for a peak corresponding to $X$ in an invariant mass 
distribution of $K \gamma \gamma$ channel which appears to have not been 
attempted with appropriate reactions.   $X$ is produced in 
reactions where kaons and pions are abundantly present.   Also since 
$X$ is an extended object, soft or relatively low momentum transfer 
reaction is needed to fuse a kaon and a pion into $X$.  Kaons and 
pions are abundantly present in relativistic heavy ion reactions.  
Despite small coalescence probability, $X$ may have been produced 
in the reactions although optimized measurement or analysis are 
probably necessary.  

We can predict properties of $X$ by its size 
which is estimated as follows.  The $K \pi$ interaction 
has a range of typically 1 fm or less.   The non-relativistic wave 
function outside the range is represented as 
\begin{equation}
\phi_{out} (r) =  N {1 \over r} exp \left( - { \sqrt{2 \mu E_B} \over \hbar c}
 r \right)
\end{equation}
where $N$ is the normalization constant, $\mu$ is a reduced mass of 
the $K \pi$ system and $E_B$ is the binding energy.   There is little 
knowledge of the wave function inside the interaction range although 
the contribution outside is dominant thus we can simply estimate 
the average radius as follows,  

\begin{equation}
<r^2> \sim \int r^2 \phi ^2_{out} (r) 4 \pi r^2 d r = \left( { (\hbar c)^2 
\over  4 \mu E_B} \right)  .
\end{equation}
For instance, $<r>$ is 4.3 fm for 5 MeV bound state and 2 fm for 30 
MeV bound state.   The present radius underestimates the real one 
since $\phi_{out}(r)$ is divergent at $r=0$ although the real wave 
function is not.   The obtained radius is fairly large compared to 
typical range of hadron 
interaction.  This is an extended object thus appropriate momentum 
transfer to excite $X$ is around 100 MeV/c or less.   This small 
momentum transfer makes production cross section of $X$ small.  The low 
momentum transfer is particularly effective to produce $X$ with small 
binding energy.


Based on the above consideration we propose here an experiment to search 
for $X$.   It is the excitation of $X$ by kaon beam irradiation on a 
proton and/or nuclear targets.   The essential point of the reaction 
is its small momentum transfer.  The momentum transfer of the 
$p(K^+, X^+)p$ reaction is shown in figure 1.  
\begin{figure}
\special{epsfile=momtr_X.eps vsize=180} 
\vspace*{7cm}
\caption{Momentum transfer of the $p(K^+,X^+)p$ reaction at 0 degrees 
is shown for $m_X=m_K + m_\pi$ and $m_X= m_K + m_\pi -30 MeV$ case.}
\end{figure}
The momentum transfer of the reaction at 0 degrees decreases gradually 
as an incident momentum $P_{K}$ increases and becomes less than 50 
MeV/c at around 2 GeV/c.  Since the spin-parity of $K^+$ and 
$X^+$ are $0^-$ and $0^+$, respectively, the angular momentum 
transfer ($\Delta \ell$) has to be 1 for the proton target.  This 
requires the momentum transfer of typically 200 MeV/c assuming 
that an interaction range is typically 1 fm.  The angular distribution 
peaks at around 10 degrees for $P_K$=1 GeV/c.  Therefore one can 
choose the incident $K^+$ momentum for experimental convenience.   
The use of nuclear target, however, may affect this choice.  For 
instance, some $J^\pi= 0^+$ nuclei like $^{16}$O have $0^-$ states 
then a $0^+$ to $0^-$ transition in the target nucleus makes a kaon 
to $X$ transition possible without angular momentum transfer.  Thus 
momentum transfer as low as 50 MeV/c directly helps to produce $X$.  
Production of $X$ through excitation of such nuclear states 
requires knowledge on a form factor which is left for future study.

$X^+$ produced by the $p(K^+, X^+)p$ reaction can be identified 
in an invariant mass spectrum obtained by measured momenta of $K^+$ 
and two $\gamma$'s.   In the present reaction there is an easier way. 
One measures $K^+$ momentum in coincidence with energetic two 
$\gamma$'s.   Only conceivable background is the reaction to produce 
$\pi^0$.   This reaction gives $K^+$ momentum similar to that of 
$X$ production.  If two $\gamma$'s are from $X$ decay, the $K^+$ 
momentum is affected by momentum carried away by two $\gamma$'s.  
On the other hand, the $K^+$ momentum is independent on two 
$\gamma$'s if they are from the $\pi^0$ decay.   The $K^+$ momentum 
is maximum when $K^+ \pi^0$ invariant mass is just at the 
threshold.  For instance, it is 0.72 GeV/c for 1 GeV/c incident $K^+$.  
On the other hand, the $K^+$ momentum can be as large as 0.96 GeV/c 
when two $\gamma$'s are detected backwards.  
The production cross section of $X$ can be scaled to the production 
of $K \pi$ at low invariant mass region.   Since $\pi^0$ production 
is clearly separated from the $X$ production by detecting two 
$\gamma$'s, abundant production of $\pi^0$ makes the search easy.  

If $X$ is proved to exist, $\Theta^+$ is likely to be the $K \pi N$
bound state.   Then the size of $\Theta^+$ can give a rough estimate 
of the width.  The decay can take place only when excitation of the 
s-wave to p-wave takes place in the $\pi N$ system and the $K- \pi N$ 
system simultaneously.  In order to realize the excitation, the three 
particles have to be present within the interaction range.   The 
typical width for the strong interaction is around a hundred MeV.  
Then the decay width is 
\begin{equation}
\Gamma \sim 100 \times \left({r_{int} \over r_\Theta}\right)^6 {\rm MeV}
\end{equation}
where $r_{int}$ is the interaction range and $r_\Theta$ is the 
radius of $\Theta^+$.  We take 1 fm as $r_{int}$ which is the 
typical interaction range for the strong interaction.  
We currently use 2 $fm$ for $r_\Theta$ where we assume that 
the binding energy of 30 MeV is carried by the pion.  The actual 
size of $\Theta^+$ should be larger since the binding energy is 
shared by three particles and weak repulsion between the kaon and 
the nucleon prefers configuration that they are apart.   
The width then becomes 
\begin{equation}
\Gamma \sim 1 {\rm MeV}
\end{equation}
This estimate can be taken as an upper limit.  We assume that if three 
particles are within $r_{int}$, the s-wave to p-wave transition takes 
place always which overestimates the width.   Also facts that $K \pi$ 
interaction range ($r_{int}$) is probably smaller than 1 $fm$ and 
$r_\Theta$ is probably larger then 2 $fm$ underestimate the real size.  
Thus the width should be narrower than 1 MeV.  

Recently reanalysis of the $K N$ scattering data was carried out.  
It gave an upper limit of 1 MeV on the width of the $K^+ n$ 
resonance \cite{arndt PS}.   This estimate is much narrower than 
the upper limit experimentally obtained and hard to accept for the 
width of resonances in such a highly excited region.   This analysis 
is consistent with our estimate of the width.  

We have discussed the possibility to explain newly observed $\Theta^+$ 
in terms of conventional pictures of the hadron physics.  We discussed 
characteristics derived from the assumption that $\Theta^+$ is 
explained as a bound state of $K \pi N$.   It is plausible that 
$u \bar{u}$ and $d \bar{d}$ combinations in the chiral soliton model 
has strong relations to the pion field.  Since mass of 
$\Theta^+$ is close to that of $K \pi N$, one has to consider the 
relation of $\Theta^+$ to a bound state of these three particles.   
This leads to the existence of the proposed $K \pi$ bound state we 
call $X$.   This possibility is 
very exotic and probably hard to believe although we show that 
current data may still allow such possibility.   We think the strongly 
attractive $K \pi$ interaction is vital to reproduce $\Theta^+$ as 
a bound state.   We present an experiment to search for $X$.   If no 
bound state is proved to exist, it is unlikely that $\Theta^+$ can be 
explained as a bound state of the three particles.   

The authors are grateful to Dr. R. Chrien for careful reading of this 
manuscript.  This work is financially supported in part by Japan 
Society for the Promotion of Science under the Japan-U.S. Cooperative 
Science Program and Japan Society for the Promotion of Science
Grant-in-Aid for Scientific Research (C) 15540275.

\bibliography{apssamp}

\end{document}